\documentclass[aps,twocolumn,groupedaddress,showpacs,floats,amssymb]{revtex4}
\usepackage[dvips]{graphicx}
\usepackage{bm}
\begin{document}

\title{Superfluid-Superfluid Phase Transitions in Two-Component Bose System}
\author{Anatoly Kuklov}
\affiliation{Department of Engineering Science and Physics,
The College of Staten Island, City University of New York,
Staten Island, New York 10314}
\author{Nikolay Prokof'ev}
\author{Boris Svistunov}

\affiliation{Department of Physics, University of
             Massachusetts, Amherst, MA 01003}
\affiliation{Russian Research Center ``Kurchatov
Institute'', 123182 Moscow }

\begin{abstract}
Depending on the Hamiltonian parameters, two-component bosons in
an optical lattice can form at least three different superfluid
phases in which both components participate in the superflow: a
(strongly interacting) mixture of two miscible superfluids
(2SF), a paired superfluid vacuum (PSF), and (at a commensurate
total filling factor) the super-counter-fluid state (SCF). We
study universal properties of the 2SF-PSF and 2SF-SCF quantum
phase transitions and show that (i) they can be mapped onto each
other, and (ii) their universality class is identical to the
$(d+1)$-dimensional normal-superfluid transition in a
single-component liquid. Finite-temperature 2SF-PSF(SCF)
transitions and the topological properties of 2SF-PSF(SCF)
interfaces are also discussed.
\end{abstract}


\maketitle

Recent remarkable success in experimental study of ultra-cold
atoms in a 3D optical lattice (OL) \cite{Greiner} signals a major
breakthrough in the field of strongly-correlated quantum lattice
systems. A simple theoretical framework which adequately describes
physics of atomic gases in OL is given by the on-site Bose-Hubbard
model \cite{Jaksh}. Its seminal prediction \cite{Fisher}---the
superfluid (SF)--Mott-insulator (MI) transition---has been
unambiguously confirmed \cite{Greiner}.

Realistic experimental perspectives of trapping several atomic
species in ultra-quantum regime have inspired theoretical studies
of multi-component systems in OL \cite{Demler,KS,Chen}. In
particular, it has been argued that the two-component commensurate
mixture of inconvertible atoms can be in the {\it
super-counter-fluid} state (SCF) \cite{KS}. In this state, the net
atomic superfluid current is impossible, and yet the equal-current
flows of two components in opposite directions are superfluid.
Another intriguing superfluid groundstate which exists in the
two-component (with equal particle numbers) Bose system in OL is
the paired superfluid vacuum (PSF) \cite{KE,Paredes}.
Qualitatively, this state is equivalent to the superfluid state of
two-atomic molecules and a BCS superconductor. An important
quantitative difference with the BCS theory is that bosonic
superfluidity exists without pairing correlations too, and PSF is
always associated with finite intra-species interaction. At the
moment, it is not clear whether PSF can be realized in atomic
gases {\it without} OL. At the 2SF-PSF transition point the
pairing interaction is necessarily strong, i.e. the scattering
length for atoms ready to form a pair is of the order of (or
larger) than the interatomic separation. Under these conditions,
metastable atomic gases are likely to become unstable from the
experimental point of view because of very large inelastic
cross-sections leading to formation of fast tight molecules (not
to be confused with loose pairs we are discussing here) and fast
atoms. In OL, this recombination channel is not an issue since now
the regime of strong/weak interaction is controlled by the ratio
of the tunnelling constant to the strength of the effective
on-site interaction, while decay rates are still controlled by the
one-site physics and are not sensitive to tunnelling.

In this Letter, we discuss universal properties of the 2SF-PSF and
2SF-SCF phase transitions. First, we prove that the two
transitions are equivalent to each other by establishing {\it
mapping} between the PSF and SCF phases. According to mapping, SCF
can be viewed as a ``molecular" superfluid, where ``molecules"
consist of particles of one component and holes of another
component. Correspondingly, the SCF transition is equivalent to
binding of two atomic superfluids into PSF. Our main focus is on
the quantum phase transition. We present strong arguments that
this transition is in the $(d+1)$-dimensional $U(1)$ universality
class, and propose an effective $(d+1)$-dimensional classical
model describing it. It allows us to relate correlation functions
in terms of the original bosonic fields to correlators of the
$U(1)$ order parameter. In the vicinity of the quantum phase
transition point, our considerations are naturally generalized to
the finite-temperature case, predicting the same $U(1)$
universality class (but now in $d$ dimensions) for the
2SF-PSF(SCF) transition at $T>0$. We verify our predictions
numerically by performing Monte Carlo simulations of a 3D
two-component closed-loop current model of Ref.~\cite{Wallin94}
which long-range critical behavior is identical to that of a
two-component 2D quantum system. Finally, we note that the
2SF-PSF(SCF) phase transition preserves the ``molecular" part of
the order parameter and discuss consequences of this fact for the
structure of vortices in hybrid systems containing 2SF-PSF(SCF)
interfaces.

The qualitative equivalence of PSF and SCF phases can be
understood on the basis of  Feynman's representation of quantum
statistics in terms of particle paths (worldlines) in imaginary
time $\tau \in [0,\beta \to \infty)$, where $\beta$ is the inverse
temperature. In this representation, the superfluid groundstate is
characterized by worldlines forming macroscopic cycles (for
brevity, we call them M-cycles), when the end of one worldline at
$\tau = \beta$ is the beginning of another worldline at time
$\tau=0$, and so on (partition function worldlines in imaginary
time are $\beta$-periodic). The qualitative difference between the
2SF and PSF groundstates is that in PSF each  A-component
worldline is bound to some B-component worldline, and in the
long-range limit there are no free single-component worldlines
forming independent M-cycles. All M-cycles are formed only by {\it
pairs} of lines, and we arrive at the picture of PSF, or molecular
superfluid. Less obvious is the fact that SCF has the same
worldline structure as PSF. The key observation is that for
integer total filing factor one may use a hole representation for
one of the components, say, component B. We readily understand
that the only worldline structure consistent with the absence of
net superfluid response is when each B-hole worldline is bound
with some A-particle worldline---this is the only possibility of
forming M-cycles out of particle-hole pairs without having
independent single-component M-cycles. Macroscopically, bound
particle-hole pairs behave like new ``molecules'' with zero
particle number charge, and their flow is equivalent to the
counter-flow of participating components.

In view of the SCF-PSF equivalence, in what follows we
discuss PSF only, implying that all results remain
valid for SCF as well.

\begin{figure}[tbp]
\includegraphics[width=6.5cm]{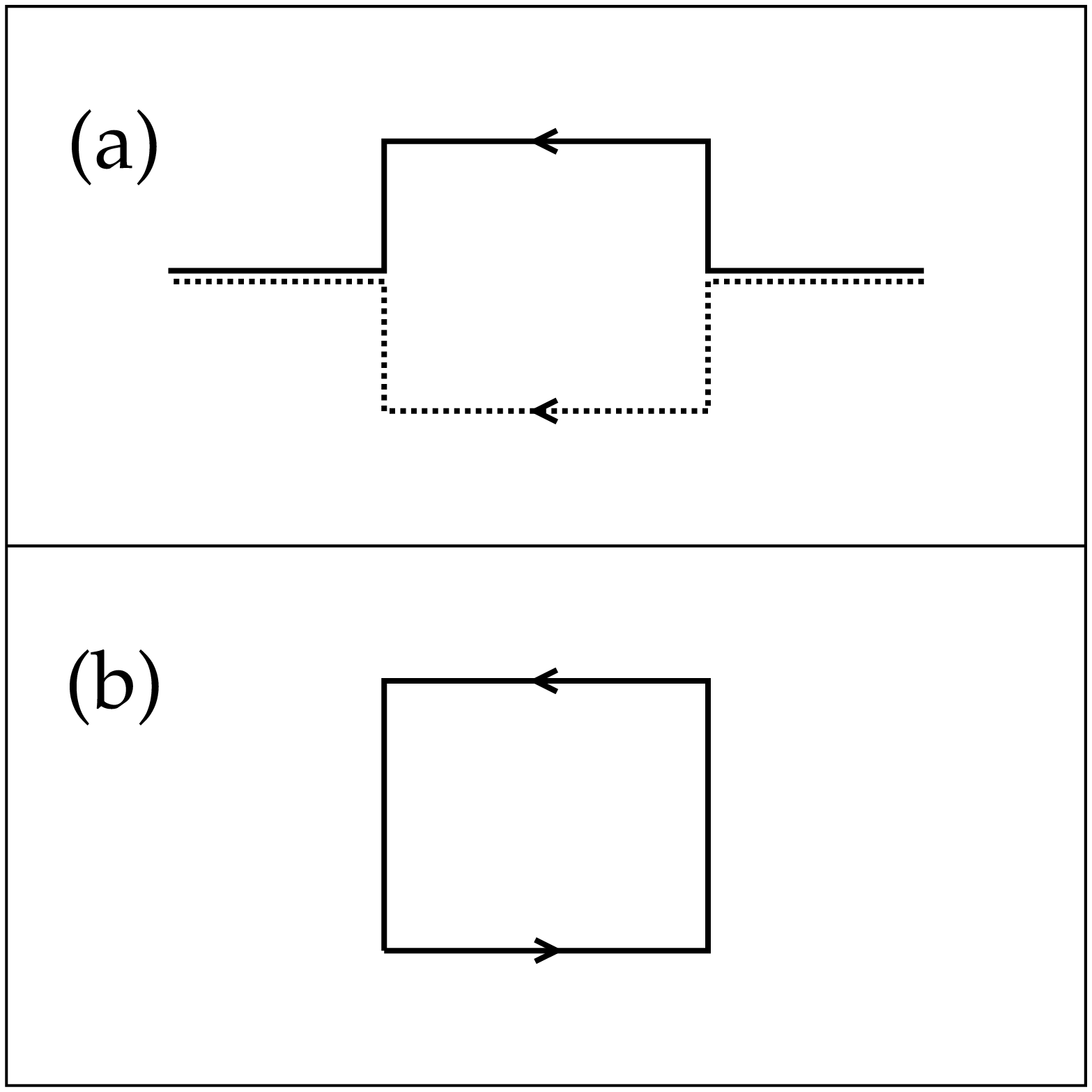}
\vspace*{-3.5cm} \caption{ (a) Unbinding fluctuations of two
coupled worldlines of the two-component system. (b) The
single-loop effective representation. Worldlines of different
species are depicted as solid and dotted lines, respectively }
\label{fig1}
\end{figure}

The worldline language presents also a ``graphic picture'' of
critical fluctuations driving the PSF-2SF transition. Suppose that
initially we are deeply in the PSF phase. Then each A-line is
closely followed by some B-line. As the coupling between
components becomes weaker, bound lines demonstrate local unbinding
fluctuations, see Fig.~\ref{fig1}(a). These fluctuations can be
viewed as single-colored {\it oriented} loops, one part of the
loop representing, say, an A-line, and the other part representing
a B-line with the reversed direction, see Fig.~\ref{fig1}(b).
Close to the critical point, unbinding loops grow larger and start
reconnecting with each other (become dense). We assume that only
large-scale loops are essential for characterizing the
universality class of the transition; the details of short-range
behavior are simply determining parameters of the critical action
for these loops. The phase transition in a system of oriented
loops in $(d+1)$ dimensions (leading to the appearance of
macroscopic-size loops) is known to describe the SF-MI transition
in a commensurate system of bosons on a $d$-dimensional lattice
(see, e.g. \cite{Wallin94}). In its turn, this transition is
equivalent to the finite-temperature phase transition between
normal and superfluid states in $(d+1)$ dimensions \cite{Fisher}.
Hence, the above reasoning suggests a mapping between the PSF-2SF
and MI-SF transitions, and establishes that the PSF-2SF transition
is in the universality class of classical $(d+1)$-dimensional
$U(1)$ models.

We now argue that in the long-wave limit our system can be also
mapped onto a $(d+1)$-dimensional model of two-color classical
rotators. This mapping is used to have one more argument in favor
of the $U(1)$ universality class, and to establish important
relations between basic correlation functions; it also provides a
natural generalization of our considerations to finite
temperatures.

The presence of lattice commensurability is not crucial for the
PSF-2SF transition since both phases involved are superfluid.
However, it proves convenient to formally assume that we are
dealing with the double-commensurate system---filling factors of
both species are integer. According to \cite{Fisher}, commensurate
$d$-dimensional lattice bosons map in the long-wave limit to a
$(d+1)$-dimensional array of rotators with the Hamiltonian
\begin{equation}
H = -\gamma \sum_{<ij>} \cos(\Phi_i - \Phi_j) \; , \label{rot_1}
\end{equation}
where $\Phi_j \in [0, 2\pi)$ is the angle of the $j$-th rotator
(or phase of the bosonic order-parameter field $\Psi(j) \sim
e^{i\Phi_j}$) and $<\cdots>$ stands for the nearest neighbor sites
on a square lattice. In our case, we need three quantum fields:
$\psi_{A}$ and $\psi_{B}$ for the two components, and the field
$\psi_{P}$ for the pairs. This suggests terms similar to
Eq.~(\ref{rot_1}) for each of the corresponding three phases.
However, one also has to account for the terms in the effective
Hamiltonian converting a bound pair into two atoms and vice versa.
In terms of the rotator model, this leads to a local term $\propto
\sum_j \cos(\Phi^{(P)}_j-\Phi^{(A)}_j-\Phi^{(B)}_j)$ reducing the
symmetry of the three-color rotor system down to $U(1)\times
U(1)$. This term introduces some (loose) constraint on the
difference between the phase $\Phi^{(P)}_j$ and the sum,
$\Phi^{(A)}_j+\Phi^{(B)}_j$. Replacing it with the rigid
constraint $\Phi^{(P)}_j=\Phi^{(A)}_j+\Phi^{(B)}_j$, we reduce the
number of independent variables from three to two---as one could
expect from the very beginning given the original $U(1) \times
U(1)$ symmetry of our system. As a result, we arrive at the
following Hamiltonian (for simplicity, we assume exchanging
symmetry between the components):
\begin{eqnarray}
H \! &=&  \! - \!\! \sum_{<ij>} (\gamma_1 \cos\Phi_{ij} +\gamma_2
\cos\Phi^{(A)}_{ij}  +\gamma_2 \cos \Phi^{(B)}_{ij}) , \; \; \; \;
\label{rot_2} \\
\Phi_j &=& \Phi^{(A)}_j + \Phi^{(B)}_j \; , \label{Phi}
\end{eqnarray}
where $\Phi_{ij}=\Phi_i - \Phi_j$ and $\Phi^{(A,B)}_{ij} =
\Phi^{(A,B)}_i- \Phi^{(A,B)}_j$. Apart from the 2SF-PSF
transition, this model can be also used to describe {\it other}
phase transitions in the {\it doubly commensurate} system, but not
otherwise.

It is convenient to introduce the phase difference
\begin{equation}
\varphi_j = \left( \Phi^{(A)}_j - \Phi^{(B)}_j \right) /2 \; ,
\label{phi}
\end{equation}
and to rewrite the Hamiltonian (\ref{rot_2}) as ($\varphi_{ij} =
\varphi_i- \varphi_j$)
\begin{eqnarray}
H =  -\sum_{<ij>} [ \gamma_1 \cos \Phi_{ij}  + 2 \gamma_2  \cos (
\Phi_{ij} /2 )
 \cos \varphi_{ij} ] \; .
\label{rot_3}
\end{eqnarray}
[The fields $\Phi$ and $\varphi$ describe charge and pseudo-spin
degrees of freedom, respectively.] Though the new variables,
$\Phi_j$ and $\varphi_j$, {\it cannot} be interpreted as angles of
new rotators---the configurational space of the original rotators
$\Phi^{(A)}_j$ and $\Phi^{(B)}_j$ is exhausted with, say,  $\Phi_j
\in [0, 2\pi)$ and $\varphi_j \in (-\pi, \pi]$, while the
Hamiltonian (\ref{rot_3}) is not $2\pi$-periodic with respect to
$\Phi_j$---for our purposes it is sufficient that just $\varphi_j$
can be viewed as a rotator angle. Indeed, in both PSF and 2SF, the
pair phase variable $\Phi_j$ is ordered and its local fluctuations
are not relevant to the criticality of the 2SF-PSF transition.
Therefore, we may simply set $\Phi_j \equiv 0$ in
Eq.~(\ref{rot_3}) which brings us to the effective one-component
rotor model for $\varphi $:
\begin{equation}
H_{\rm 2SF-PSF} = -2 \gamma_2 \sum_{<ij>} \cos\varphi_{ij} \; .
\label{XY}
\end{equation}
The transition thus is the superfluid--normal-fluid transition in
the $\varphi$-channel (which means localization in the pseudo-spin
sector); the corresponding complex order parameter is $\psi({\bf
X}) \propto e^{i\varphi({\bf X})}$, where ${\bf X}$ is the
space-time radius-vector treated as a continuous variable in the
long-wave limit. Given this order parameter and Eq.~(\ref{phi})
relating $\varphi$ to the original phases ($\Phi^{(A)} = -
\Phi^{(B)} = \varphi$), we immediately find the critical behavior
of various correlation functions
\begin{eqnarray}
\langle \psi^{\dagger}_{\rm A}(X) \psi^{}_{\rm A}(0)\rangle \sim
\langle \psi^{\dagger}_{\rm B}(0) \psi^{}_{\rm B}(X)\rangle
\sim \langle \psi^{}_{\rm A}(0) \psi^{}_{\rm B}(X)\rangle \nonumber \\
 \sim \langle \psi^{\dagger}(X) \psi(0)\rangle \; .
\label{corr}
\end{eqnarray}

Now we note that  $(d+1)$-dimensional model (\ref{XY}) with large
but finite size in the $\tau$-direction describes  the initial
part of the finite-temperature 2SF-PSF line in the vicinity of the
quantum critical point.  We thus establish the universality
class---$U(1)$ in $d$ dimensions---for the finite-temperature
second-order 2SF-PSF transition. Since the order parameter for the
transition is $\sim e^{i\varphi}$ (``molecular" order parameter is
not critical), we arrive to a rather counter-intuitive conclusion
that with increasing temperature the transition is {\it from} 2SF
{\it to} PSF. [Clearly, relations (\ref{corr}) take place on the
finite-temperature critical line as well.]

This finite-temperature 2SF-PSF transition survives even when the
two components have slightly different densities and the
groundstate is {\it inevitably } 2SF (both $\langle \psi_A \rangle$ and
$\langle \psi_B \rangle$ are non-zero). Away from the quantum
critical point, the 2SF-PSF transition can be viewed as the
superfluid to normal fluid transition in the (dilute) sub-system
of excessive particles.

Equation (\ref{rot_3}) is also useful for understanding the
structure of vortices across the interface between the PSF and 2SF
phases. Experimentally, interfaces naturally arise in
inhomogeneous systems (in a confining potential, particle
densities drops to zero at the boundary, and, e.g. the SCF phase,
which requires commensurability, may not survive at the edge).
Suppose one creates a vortex in a PSF phase and then adiabatically
removes OL and the trapping potential to observe the system by the
standard technique of absorption imaging \cite{imaging}. When the
lattice potential is turned off, the system will behave as two
weakly interacting gases. The question now is: Do vortices in the
PSF phase transform (and how) into vortices in the resulting
weakly interacting system \cite{Holland}?

\begin{figure}[tbp]
\includegraphics[width=6.5cm]{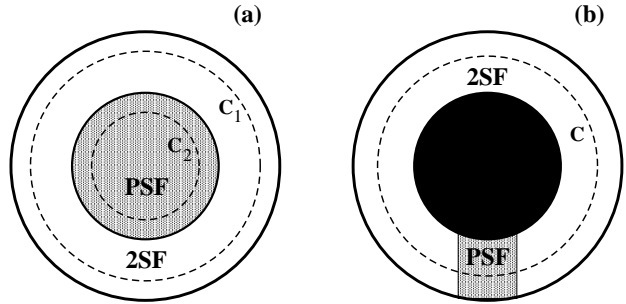}
\vspace*{-4.cm} \caption{Different geometries of PSF-2SF
interfaces. (See discussion in the text.)} \label{fig2}
\end{figure}
System inhomogeneity implies that at intermediate stages
of the potential turning off, there will be an interface
similar to one shown in Fig.~\ref{fig2}(a).
Ultimately, the interface shrinks and disappears with the PSF phase,
while the topological structure of the 2SF state remains the same
as it was when the interface existed.
To understand this structure we resort to the rotator model (\ref{rot_3}).
In both PSF and 2SF the phase field $\Phi$ is ordered and
thus the circulation of $\nabla \Phi$ does not change across the interface.
Since the phase field $\varphi$ is disordered inside the PSF phase,
there are no topological constraints associated with it. Taking into account
(\ref{Phi}), we arrive at the following rule for topological
charges:
\begin{equation}
I^{(A)} + I^{(B)} = I \; , \label{rule}
\end{equation}
where $I^{(A,B)}$ and $I$ are {\it integer} topological charges in
2SF and PSF correspondingly
\begin{equation}
I^{(A,B)} ={1 \over 2\pi} \oint_{C_2} \! \! \nabla \Phi^{(A,B)} \,
{\rm d}{\bf l} \, , ~~ I = {1 \over 2\pi} \oint_{C_1} \! \!
\nabla \Phi \, {\rm d}{\bf l} \;. \label{C}
\end{equation}
We see that vortices in PSF always induce vortices in the 2SF
phase fields $\Phi^{(A)}$ or/and $\Phi^{(B)}$, and thus makes it
possible to observe the circumstantial evidence of the PSF vortex
in the absorption image of the weakly interacting 2SF state.
However, the values of topological charges are not unambiguously
defined. For example, if $I=1$ then $(I^{(A)}=1, I^{(B)}=0)$ and
$(I^{(A)}=0, I^{(B)}=1)$, are consistent with Eq.~(\ref{rule}), as
well as, say, $(I^{(A)}=2, I^{(B)}=-1)$. This implies that
particular values of $I^{(A)}$ and $I^{(B)}$ will depend on
details of the experimental setup determining the lowest
energy configuration. For example,
if the two components have
different superfluid stiffnesses and, initially, there was
one vortex in the PSF, then, after creating the interface and removing
the PSF, the vortex will reside in the component with lower
stiffness.

Another interesting geometry is shown in Fig.~\ref{fig2}(b). Using
arguments identical to those presented above, we see that the only
integer topological charge on contour $C$ is $I$. While the sum of
integrals for $I^{(A)}$ and $I^{(B)}$ still equals $I$, separately
they are ill defined on $C$, because the phase ~$\varphi$
experiences large zero-point fluctuations in the PSF region.
Suppose, then, that initially there were no PSF phase at all, and
the topological charges of components were, say, $I^{(A)}=1$ and
$I^{(B)}=0$. Imposing OL to create PSF will eliminate quantization
for the individual phases $\Phi^{(A,B)}$, while preserving the sum
$I^{(A)} + I^{(B)}=1$. Accordingly---since no memory about the
initial values $I^{(B)}, I^{(A)}$ is retained---further removal of
the OL may result in the final 2SF state with $I^{(A)}=0,
I^{(B)}=1$. Similarly to the previous setup, if the two components
have different superfluid stiffnesses, then, after the cycle of
switching on and off OL, the circulation will reside in the
component with lower stiffness.

There are several options to verify above considerations
numerically. One is a direct simulation of the two-component
$d$-dimensional Bose-Hubbard Hamiltonian at very low temperature.
However, the universality class of the phase transition and the
relevant long-wave description of critical fluctuations, may be
also obtained for  the $(d+1)$-dimensional classical lattice model
built on particle trajectories in discrete imaginary time. One of
the quantum-to-classical mappings for the Bose-Hubbard
Hamiltonian---the $J$-current model---was developed in
Ref.~\cite{Wallin94}, and we straightforwardly generalized it to
the two-component case. Our choice to simulate the classical
action was motivated only by reasons of numeric efficiency.
Recently developed quantum and classical Worm Algorithms do not
suffer from critical slowing down \cite{worm2001}, but the
classical one is superior because of its simplicity (it was
already successfully applied to the disordered one-component
$J$-current  model \cite{AS}). We defer details of simulations
performed for the $d+1=3$ case to a longer paper \cite{preprint}
and simply mention here results. The correlation radius and the
correlation function exponents for the 2SF-PSF transition agreed
with the known values for the 3D U(1)-universality class \cite{nu}
within $1\div 2$\% accuracy. We have directly verified that
non-trivial relations between the correlation functions given by
Eq.~(\ref{corr}) hold true at the critical point, and deviations
from Eq.~(\ref{corr}) are barely visible even at distances as
small as 5 lattice constants.  We have observed the qualitative
prediction of the model (\ref{XY}) about the transition from 2SF
to PSF with increasing temperature.

Summarizing, we have shown that two strongly-correlated phases of
the two-component bosonic system---the superfluid state of pairs
and the counter-flow superfluid---are macroscopically equivalent.
We have presented arguments supported by results of numeric
simulations, that quantum phase transitions leading to formation
of these phases from the state of two miscible superfluids are in
the universality class of superfluid--Mott-insulator transition in
a single-component bosonic system. The finite-temperature
2SF-PSF(SCF) transitions are in the universality class of a
single-component superfluid--normal-fluid transition. The proposed
two-color rotator model correctly describes the critical behavior
of various correlators and---in the spatially inhomogeneous
case---yields a simple rule for ``sewing'' topological defects
across the boundary between the phases. On the basis of this rule
it is possible to observe the evidence of the 2SF-PSF(SCF) phase
transitions even without directly detecting it.

\end{document}